\documentclass[pra, superscriptaddress, aps,showpacs,twocolumn]{revtex4}

\usepackage{graphicx}
\usepackage{bm}

\usepackage[usenames]{color}

\begin{document}

\title{Quantum teleportation and entanglement distribution over 100-kilometre free-space channels}

%\author{authors}
%\iffalse
\author{Juan Yin}
\thanks{These authors contributed equally to this work}
\author{Ji-Gang Ren}
\thanks{These authors contributed equally to this work}
\author{He Lu}
\thanks{These authors contributed equally to this work}
\author{Yuan Cao}
\author{Hai-Lin Yong}
\author{Yu-Ping Wu}
\author{Chang Liu}
\author{Sheng-Kai Liao}
\author{Fei Zhou}
\author{Yan Jiang}
\author{Xin-Dong Cai}
\author{Ping Xu}
\author{Ge-Sheng Pan} 
\affiliation{Shanghai Branch, National Laboratory for Physical Sciences at Microscale and Department of Modern Physics, University of Science and Technology of China, Shanghai 201315, China}
\author{Jian-Jun Jia}
\affiliation {Shanghai Institute of Technical Physics, Chinese Academy of Sciences, Shanghai 200083 China}
\author{Yong-Mei Huang}
\affiliation {The Institute of Optics and Electronics, Chinese Academy of Sciences, Chengdu 610209, China}
\author{Hao Yin}
\affiliation{Shanghai Branch, National Laboratory for Physical Sciences at Microscale and Department of Modern Physics, University of Science and Technology of China, Shanghai 201315, China}
\author{Jian-Yu Wang} 
\affiliation {Shanghai Institute of Technical Physics, Chinese Academy of Sciences, Shanghai 200083 China}
\author{Yu-Ao Chen} 
\author{Cheng-Zhi Peng}
\author{Jian-Wei Pan}
\affiliation{Shanghai Branch, National Laboratory for Physical Sciences at Microscale and Department of Modern Physics, University of Science and Technology of China, Shanghai 201315, China}

\date{\today}

\begin{abstract}
A long standing goal for quantum communication is to transfer a quantum state over arbitrary distances. Free-space quantum communication provides a promising solution towards this challenging goal. Here, through a 97-km free space channel, we demonstrate long distance quantum teleportation over a 35-53~dB loss one-link channel, and entanglement distribution over a 66-85~dB high-loss two-link channel.  We achieve an average fidelity of {80.4(9)}~\% for teleporting six distinct initial states and observe the violation of the Clauser-Horne-Shimony-Holt inequality after distributing entanglement. Besides being of fundamental interest, our result represents a significant step towards a global quantum network. Moreover, the high-frequency and high-accuracy acquiring, pointing and tracking technique developed in our experiment provides an essential tool for future satellite-based quantum communication.
\end{abstract}

\maketitle

%\pacs{}

%\maketitle

The concept of quantum entanglement, a fundamental feature of quantum mechanics, lies at the heart of quantum information processing\cite{Pan12}. With the help of quantum entanglement among distant locations, quantum communication (QC) can be achieved over arbitrary distances by quantum teleportation and entanglement distribution. The discovery of the former\cite{Bennett93} has led to seminal quantum information processing protocols like large-scale QC\cite{Briegel98}, and the latter has become an indispensable process in many QC schemes such as  entanglement based quantum key distribution\cite{Ekert91} and test of non-locality of quantum mechanics\cite{Bell64,Freedman72,Aspect81,Weihs98}.

Free-space optical quantum channel appears to be very promising to implement long-distance QC. As information carrier photons are fast and robust. 
Since the photon loss is almost negligible in the outer space and the effective thickness of the atmosphere is only about10~km, free-space optical channel, first used for quantum key distribution\cite{Hughes02,Kurtsiefer02}, can be far superior over fiber links and is most promising for satellite-based QC on a global scale.

Quantum teleportation was first demonstrated with entangled photons\cite{Bouwmeester97} in 1997. Later, various developments have been achieved in laboratory, including the demonstration of entanglement swapping\cite{Pan98}, open-destination teleportation\cite{Zhao04} and teleportation of two-bit composite system\cite{Qiang06Teleportation}. Entanglement distribution has been shown with fiber links\cite{Marcikic04,Huebel07,Zhang08,Dynes09}. In addition, ``practical'' quantum teleportation have been realized via fiber links\cite{Marcikic03,Ursin04} and limited to a distance of about one kilometer. 
Experiments  have achieved free space distribution of entangled photon pairs over distances of 600~m\cite{Aspelmeyer03} and 13~km\cite{Peng05}. Later, entangled photons were transmitted over 144~km\cite{Ursin07,Fedrizzi09}. In these experiments,  either only one photon was transmitted\cite{Ursin07} or the entangled photon pair was transferred together\cite{Fedrizzi09} using only a one-link channel. 
Most recently,  following a modified scheme\cite{Boschi98}, quantum teleportation over a 16 km free-space links was demonstrated\cite{Jin10} with a single pair of entangled photons. However in this experiment, the unknown quantum state must be prepared on one of the resource entangled qubits and therefore cannot be presented independently.

In our experiment, we first demonstrate quantum teleportation of an independent unknown state  through a 97-km one-link optical free-space channel with multi-photon entanglement, following the original scheme\cite{Bennett93, Bouwmeester97}. An average fidelity of 80.4(9) \% can be achieved for six distinct initial states over a 35-53~dB loss quantum channel. Furthermore we report an entanglement distribution over a  two-link quantum channel, where the entangled photon pairs are separated by 101.8~km and the total channel loss is up to 66-85~dB. Violation of the Clauser-Horne-Shimony-Holt (CHSH) inequality\cite{CHSH} is successfully observed. The successful quantum teleportation and the entanglement distribution over such channel losses in combination with our high-frequency and high-accuracy acquiring, pointing and tracking (APT) technique show the feasibility of satellite-based ultra-long-distance quantum teleportation and global-scale QC.  

\section*{Teleportation through one-link channel}

\begin{figure}[t!]
\includegraphics[width=0.95\linewidth]{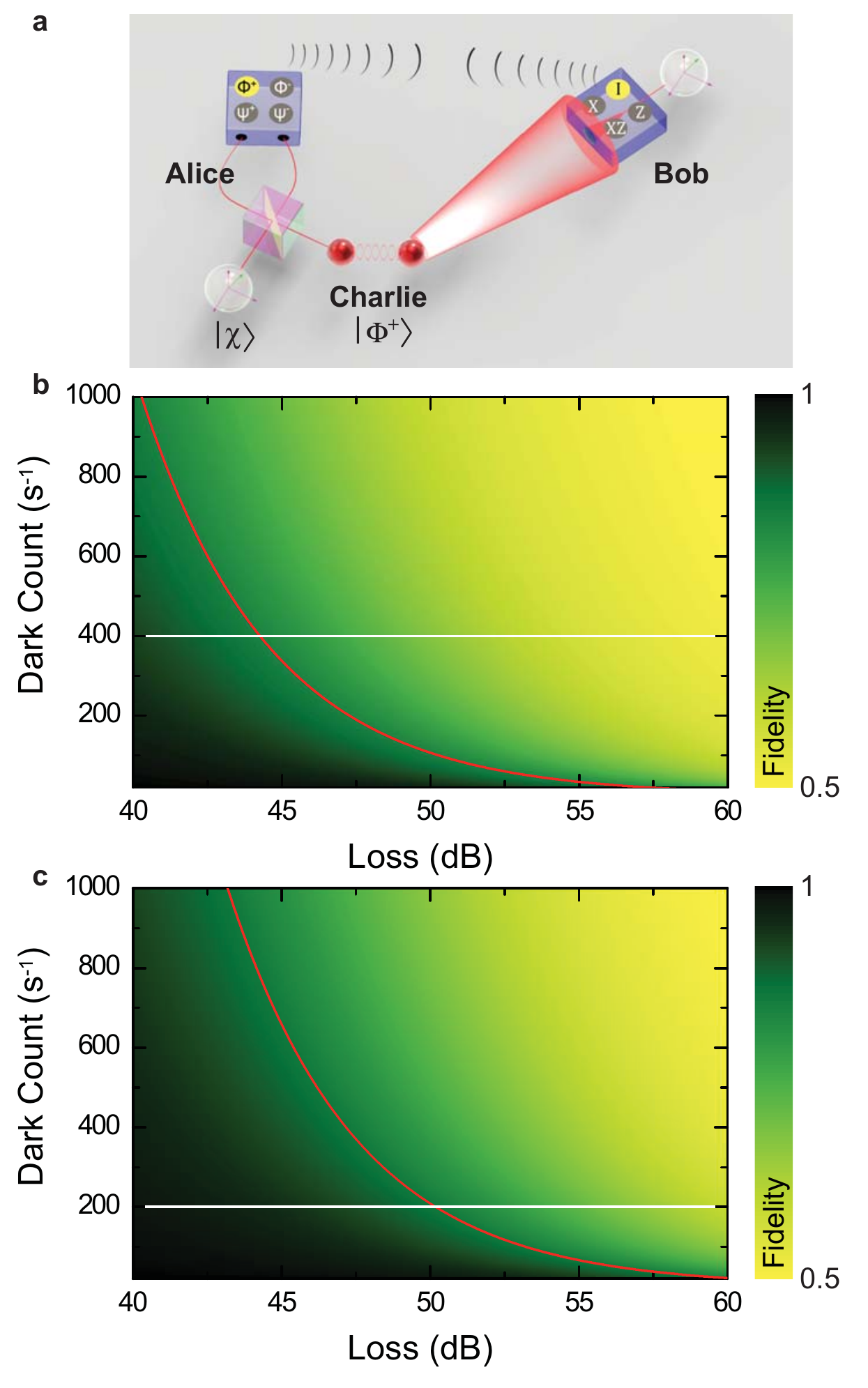}
\caption{Experimental scheme and calculated fidelity for practical free-space quantum teleportation.  \textbf{a} Schematic drawing for practical free-space quantum teleportation. Charlie distributes an entangled pair of photons 2 and 3 to Alice and Bob, where Bob is at a distant location. Due to the finite size of the telescopes Bob and Charlie can use and the diffraction limit, Bob will receive the signal photon with very high loss. Alice then performs a joint Bell-state measurement (BSM) on the initial particle and one of the entangled photons from Charlie, projecting them onto an entangled state. After she has sent the result of her measurement as classical information to Bob, he can perform a unitary transformation (U) on his photon to obtain the initial state. Note that in our experiment, the unitary transformation was not applied. \textbf{b} Calculated fidelity for ideal optics and an entangled photon source with the generation probability of 0.083 pair per pulse and overall detection efficiency of 12.3\% used in Ref. \cite{Chaoyang07} as a function of total channel loss and the dark count on Bob's side. The red line shows the classical limit\cite{Popescu94} of 2/3. The white line shows the lowest dark count rate currently achieved via free-space link\cite{Fedrizzi09}. \textbf{c} Calculated fidelity for ideal optics for our ultra-bright entangled photon source with the generation probability of 0.1 pair per pulse and  overall detection efficiency of 23.6\%. %
\label{Fig_Scheme}}
\end{figure}

A schematic illustration of free-space quantum teleportation is shown in Fig.~\ref{Fig_Scheme}a. Alice has a photon in an unknown quantum state
$|\chi\rangle_1=\alpha|H\rangle+\beta|V\rangle$, where $H~(V)$ represents the horizontal (vertical) polarization. %where $\alpha$ and $\beta$ are two complex numbers satisfying  $|\alpha|^2+|\beta|^2=1$, 
Alice wants to transfer the photon to Bob, who is at a distant location. To do so, Charlie first distributes an entangled photon pair 2 and 3, $|\Phi^{+}\rangle_{23}=(|H\rangle|H\rangle + |V\rangle|V\rangle)/\sqrt{2}$, to Alice and Bob, respectively. The combinative state of the three photons can be rewritten as
\begin{eqnarray}
  |\chi\rangle_1&\otimes&|\Phi^{+}\rangle_{23} =\frac{1}{2}\left(|\Phi^{+}\rangle_{12}|\chi\rangle_{3}+|\Phi^{-}\rangle_{12}(Z|\chi\rangle_{3}) +\right. \nonumber \\
 &&  \left.|\Psi^{+}\rangle_{12}(X|\chi\rangle_{3})+|\Psi^{-}\rangle_{12}(XZ|\chi\rangle_{3})\right),
 \label{eq:tele}
\end{eqnarray}
where  $|\Phi^{\pm}\rangle_{12}=\left (|H\rangle_1|H\rangle_2\pm|V\rangle_1|V\rangle_2\right)/\sqrt 2$, $|\Psi^{\pm}\rangle_{12}=\left (|H\rangle_1|V\rangle_2\pm|V\rangle_1|H\rangle_2\right)/\sqrt 2$ are the four Bell states, and Z, X are Pauli operators ,which act as unitary transformations.
Then, if Alice performs a joint Bell-state measurement (BSM) on her two photons, photon 3 is instantaneously projected into the four states $|\chi_{3}\rangle$, $Z|\chi_{3}\rangle$, $X|\chi_{3}\rangle$ and $XZ|\chi_{3}\rangle$, respectively. Thus, after receiving the BSM result from Alice via a classical channel, Bob can apply the appropriate unitary transformation to convert the state of photon 3 to the initial state\cite{Bennett93, Bouwmeester97}.

Practically, the quantum channel between Charlie and Bob always has losses due to the finite size of the telescopes on both sides (see Fig.~\ref{Fig_Scheme}a). Typically, for an uplink of the ground station to a satellite,  this loss can be up to 45 dB (calculated for 20~cm satellite optics at an orbit height of 500~km). The losses themselves only reduce the success probability of the teleportation. With perfect detectors and without background light, the channel losses would not be the limiting factor. However, because of the intrinsic dark counts by the detector and additional background from the environment, there is a chance that a dark count produces an error each time a photon is lost. When the probability of a dark count becomes comparable to the probability that a photon is correctly detected, the signal-to-noise ratio tends to 0. In order to overcome this  problem, one has two possibilities: increasing the brightness of the entangled photon source or reducing the dark count rate. The calculated fidelity as a function of dark count rate and channel loss is shown in Fig.~\ref{Fig_Scheme}b. One can see that even with ideal optics, it is not possible to demonstrate successful quantum teleportation under a 45 dB loss channel with the currently achieved lowest dark count rate\cite{Fedrizzi09} of $\sim$400~s$^{-1}$.

\begin{figure*}[t!]
\begin{center}
\includegraphics[width=0.95\linewidth]{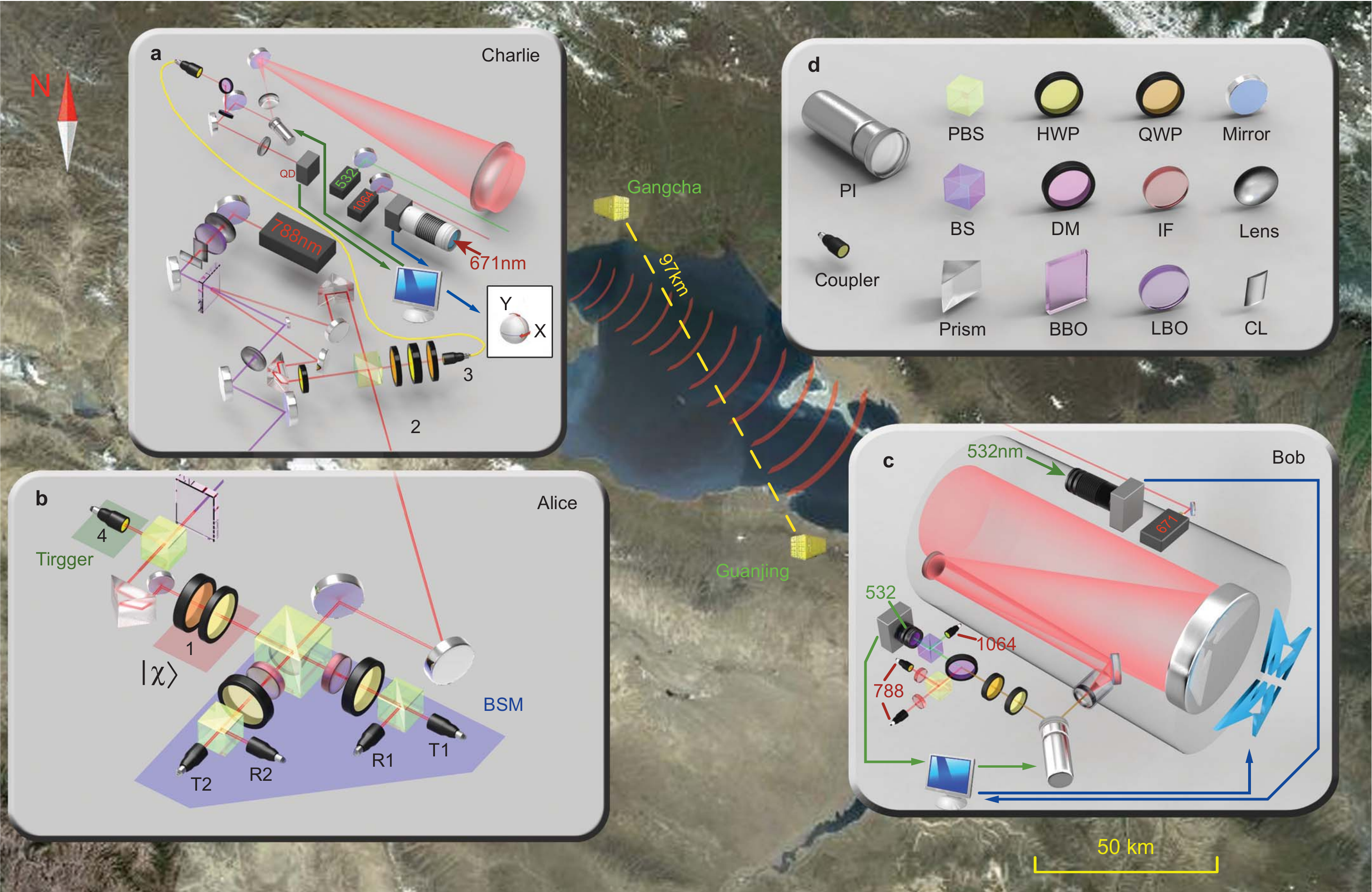}
\caption{Bird's-eye view and schematic diagram for free-space quantum teleportation. \textbf{a}, Entanglement generation and distribution on Charlie's side. A near infrared pulse (788~nm) is focused on an LBO crystal to create an ultraviolet laser pulse, which is then focused with two cylindrical lenses (CL)  and passed through a 2 mm nonlinear BBO crystal. By an SPDC process, an entangled photon pair is  created. An interferometric Bell-state synthesizer is utilized  to disentangle the temporal from the polarization information\cite{Yao11}. While photon 2 is then directly sent to Alice for a BSM, photon 3 is guided to a refractor telescope through a fiber and sent to Bob. An HWP sandwiched between two QWPs constitute the fiber polarization compensation. Coaxial with the telescope, there is a green laser (532 nm) for system tracking and a infrared laser (1064 nm) for synchronization. The green arrows indicate the fine tracking system which consists of a four-quadrant detector (QD) and a fast steering mirror driven by piezo ceramics (PI). The blue arrows indicate the coarse tracking system which consists of a wide-angle camera and a two-dimensional rotatable platform. \textbf{b}, Initial state preparation and BSM on Alice's side.
Alice sends the UV laser through a collinear BBO, creating another photon pair which is later separated in path by a PBS. An HWP and a QWP is applied in path 1 to prepare the initial unknown quantum state to be teleported. Alice interferes the initial state 1 and the photon 2 from Charlie using another PBS. A 22.5$^{\circ}$ HWP and PBS is placed at both outputs for polarization analysis. A coincidence between detectors T1 and T2 (R2) or R1 and R2 (T2) indicates the incident state of $|\Phi^{+}\rangle$ ($|\Phi^{-}\rangle$).
\textbf{c}, Polarization analysis on Bob's side.
Bob receives photon 3 with a 400mm diameter off-axis reflecting telescope. A polarization analyzer is assembled at the telescope's exit, containing an HWP, a QWP, a PBS, and two multi-mode fiber-coupled SPCMs. Coaxial with the receiving telescope, there is another high-power beacon laser (671 nm) for system tracking. The blue and green arrows indicate the coarse and fine tracking system, respectively.
\textbf{d}, Symbols used for the setup.
\label{Fig_Setup}}
\end{center}
\end{figure*}

Experimentally, we start with an ultra-bright entangled photon source\cite{Yao11} based on type-II spontaneous parametric down-conversion\cite{Kwiat95} (SPDC). As shown in Fig.~\ref{Fig_Setup}a, on Charlie's side (located at Gangcha next to Qinghai Lake, 37$^{\circ}$16$^{\prime}$42.41$^{\prime\prime}$ N 99$^{\circ}$52$^{\prime}$59.88$^{\prime\prime}$ E, altitude 3262~m), a  femtosecond ultraviolet (UV) laser is created by frequency doubling a pulsed laser (MIRA, central wavelength of 788~nm with a duration of 130~fs and a repetition rate of 76~MHz) with an LBO crystal (LiB$_3$O$_5$). The UV laser is further guided to pump a noncollinear type-II $\beta$ -barium (BBO) crystal, resulting in  a pair of polarization entangled photons in the state $|\Psi\rangle=\left ( |H_oV_e\rangle+|V_eH_o\rangle\right)/\sqrt 2$ with temporal and polarization information also entangled\cite{Yao11}, where $o$ and $e$ indicate the polarization with respect to the pump. With an interferometric Bell-state synthesizer\cite{Yao11,Kim03},  we disentangle the temporal from the polarization information by guiding photons of different bandwidths through separate paths, resulting in the  desired entangled photon source  $|\Phi^{+}\rangle_{23}=\left(|HH\rangle+|VV\rangle\right)_{23}/\sqrt 2$. Charlie then distributes the two photons 2 and 3 to Alice and Bob, respectively.

To prepare the unknown quantum state to be teleported, Alice uses the UV laser to pump a collinear BBO crystal which emits photons along the pumping light direction (see Fig.~\ref{Fig_Setup}b). The generated photons are $|HV\rangle_{14}$, which are then split by a PBS after the pumping laser has been filtered out. A half wave plate (HWP) and a quarter wave plate (QWP) are applied in path 1 to create the initial state. Under a trigger on path 4, Alice creates the state she wants to teleport. Then, Alice performs a joint BSM  on photon 1 and 2 by interfering them on a PBS and performing polarization analysis on the two outputs. The subsequent coincidence measurements can identify the $|\Phi^{\pm}\rangle$ Bell states in our experiment.

In the experiment, the power of the pulsed UV laser was about 1.3~W after frequency doubling. The observed average two-fold coincidence rate  was  4.4$\times$10$^5$~s$^{-1}$ with 3~nm filters in the $e$-ray path 2 and 8~nm filters in the $o$-ray path 3. The visibility of the entangled photon pair was about 91\% in the $\vert H\rangle/\vert V\rangle$ basis and 90 \% in the $\vert +\rangle/\vert-\rangle$ basis, where $|\pm\rangle=1/\sqrt{2}(|H\rangle\pm|V\rangle$.  The generation rate of the entangled photons was about 0.1 pair per pulse, and the overall detection efficiency was 0.236 locally. The count rate of photon pair $|HV\rangle_{14}$ generated by the collinear BBO crystal was 6.5$\times$10$^5$~s$^{-1}$, with the photon in path 1 ($e$-ray) filtered by an interference filter ($\Delta_{\tiny{\mbox{FWHM}}}=$3~nm) and the photon in path 4 ($o$-ray) detected without filters.
In the joint BSM, the observed visibility of interference for the photons overlapping on the PBS was 0.6. Finally, we observed about a 2$\times$10$^3$~s$^{-1}$ counting rate for four-fold coincidence locally. As comparison,  we plot the calculated fidelity as a function of dark count rate and channel loss for our source (Fig.~\ref{Fig_Scheme}c). 

On the other hand, Charlie sends photon 3 with a compact transmitting system to Bob on the other side of Qinghai Lake, as shown in Fig.~\ref{Fig_Setup}a. A 127mm f/7.5 extra-low dispersion alternative public offering refractor telescope is employed as an optical transmitting antenna. By minimizing the color dispersion, we obtain superior sharpness and color correction. For near-diffraction-limited far-field divergence angles, we design systems to substantially reduce chromatic and spherical aberrations. Finally, the divergence angle of our compact quantum transmitter is about 20~$\mu$rad. 

As shown in Fig.~\ref{Fig_Setup}c, Bob (GuanJing 36$^{\circ}$32$^{\prime}$43.31$^{\prime\prime}$N 100$^{\circ}$28$^{\prime}$9.81$^{\prime\prime}$E, altitude 3682~m) receives photon 3 with a 400mm diameter off-axis reflecting telescope. 
%By selecting the appropriate eyepiece, the telescope gets spot compression 100 times. 
An integrated measurement system, consisting of an HWP, a QWP and a PBS,  is assembled at the telescope's exit to measure any arbitrary state. Then, the photons are coupled in multi-mode fibers by a non-spherical lens. By selecting the appropriate fiber core and focal length, we compress the receiver field of view to 70~$\mu$rad, which directly improves the system's signal-to-noise ratio. In front of the non-spherical lens, two band-pass filters ($\Delta_{\tiny{\mbox{FWHM}}}=80$~nm) and one narrow-band interference filter ($\Delta_{\tiny{\mbox{FWHM}}}=10$~nm) are used to reduce background noises (IF shown in Fig.~\ref{Fig_Setup}c).  Finally the photons are detected by the single-photon counting modules (SPCM) with ultra-low dark counts ($<$ 20~s$^{-1}$). The noise that we observed, including the dark counts and ambient counts, is in total about 160~s$^{-1}$ to 300~s$^{-1}$. The noise mainly depends on the position of the moon and averagely we obtain about 200~s$^{-1}$. 

\section*{Acquiring, pointing and tracking system}

\begin{figure}[t!]
\begin{center}
\includegraphics[width=0.95\linewidth]{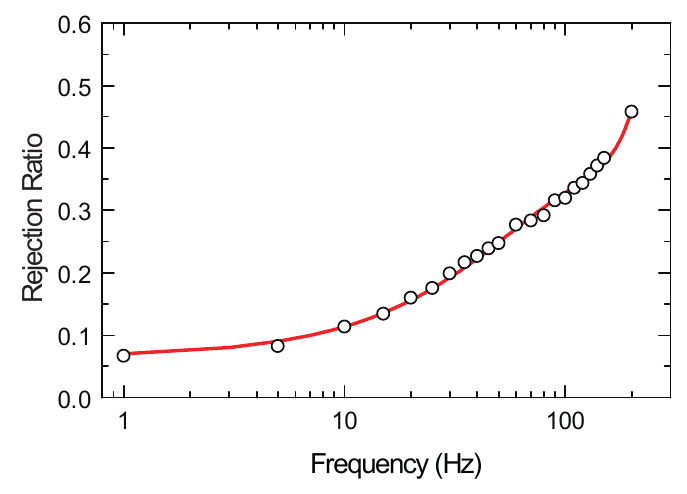}
\caption{The rejection ratio values as a function of the perturbation frequency. We perform this test indoors. An additional PI-FSM driven by a frequency-adjustable sinusoidal signal is used to induce the perturbation into fine tracking system of the transmitter. Here we define the ratio of the amplitudes between fine tracking on and off as rejection ratio. The solid line is a polynomial fit as the guide for eyes. The close-loop bandwidth of the fine tracking is defined as associated  perturbation frequency at a rejection ratio  of 0.5 (-3dB). It is clear that the bandwidth is more than 150 Hz.
\label{Fig_APT}}
\end{center}
\end{figure}

In addition to the optical design previously mentioned, we also equip an APT system to account for effects due to ground settlement, mechanical deformation, atmospheric turbulence, etc. As shown in Fig.~\ref{Fig_Setup}a, on the sender Charlie's side, coaxial with the entangled photon 3, there is a continuous green laser (532~nm, 200~mw, 1.5~mrad) for system tracking and a pulsed infrared laser (1064~nm, 10~kHz, 50~mw, 200~$\mu$rad) for synchronization. On the receiver Bob's side, coaxial with the receiving telescope, there is a high-power beacon laser (671~nm, 2~w, 200~$\mu$rad) for system tracking (Fig.~\ref{Fig_Setup}c). When the optical link was established for the first time, Charlie achieved acquiring by Global Positioning System (GPS) coordinates and light guide. At the same time, he fired the beacon light (532~nm) pointing to the receiver Bob. Bob then achieved acquisition and fired another beacon light (671~nm) pointing back to Charlie. 

The tracking system is composed by a cascade close-loop control system (the blue and green arrows in Fig.~\ref{Fig_Setup}a and \ref{Fig_Setup}c). On Charlie's side, the beacon laser from the receiver Bob is detected by a wide-angle camera. With a feedback loop, the coarse alignment of the entire optical system is achieved by the two-dimensional rotatable platform in both azimuth and elevation (blue arrows in Fig.~\ref{Fig_Setup}a). Similarly, the fine tracking indicated by the green arrows is achieved by the four-quadrant detector (QD) and fast steering mirror driven by piezo ceramics. Furthermore, the fine tracking system shares the same optical path as the quantum channel and is later separated by a dichroic mirror (DM). Thus, a much higher tracking accuracy can be obtained. The closed-loop bandwidth of the fine tracking is more than 150~Hz (see the inset in Fig.~\ref{Fig_APT}), which is sufficient to overcome most of the atmospheric turbulence\cite{Strohbehn78}. Finally, with this system design the tracking accuracy is better than 3.5 $\mu$rad over the 97km free-space link.

As indicated by the blue and green arrows in Fig.~\ref{Fig_Setup}c, there are also coarse and fine tracking on the receiver Bob's side, by closed-loop control via the telescope's own rack and piezo ceramics. Since the main purpose of the tracking system at the receiver is to reduce the low frequency shaking due to ground settlement and passing vehicles, the closed loop bandwidth is about 10~Hz. The APT system is designed for tracking an arbitrarily moving object, which can be directly utilized for a satellite-based QC experiment. In experiments between fixed locations, the first two steps, acquiring and pointing, do not need to be done every day. 

In addition, we utilize a wireless Bridge for data transmission and classical communication between Alice and Bob. A high-accuracy Time-to-Digital Converter is used to  independently record the arrival time of signals at both Alice's and Bob's station. A pulse per second produced by the GPS are added to synchronize the starting time.  With the help of the pulsed synchronization laser, we achieve a time synchronization accuracy of better than 1~ns (see Supplementary Information for details). 

\section*{Results for quantum teleportation}

\begin{table}
\caption{Fidelity of quantum teleportation over 97~km. The data collection was accumulated for 14400~s. The errors denote the statistical error, which is $\pm$ 1 standard deviation. 
}
\centering
\arraycolsep=9pt
\renewcommand{\arraystretch}{1}
\begin{tabular}{c | c}
  \hline
  % after \\: \hline or \cline{col1-col2} \cline{col3-col4} ...
  State & Fidelity \\\hline\hline
  $H$ & 0.814$\pm$0.031 \\ 
  $V$ & 0.886$\pm$0.024 \\
  $+$ & 0.773$\pm$0.031 \\
  $-$ &  0.781$\pm$0.031 \\
  $R$ & 0.808$\pm$0.026 \\
  $L$ &  0.760$\pm$0.027 \\
  \hline
\end{tabular}
\label{Tab:100kmtele:Result0}
\end{table}

After debugging the entire system, we measured the channel loss in the Qinghai Lake district over 97~km horizontal atmospheric transmission at near ground levels.
%The far-field spot size was 3.5--7~m, depending on weather conditions, and correspondingly, the optimal loss was 35--53~dB. 
The measured link efficiency was between 35 and 53~dB, in which 8~dB was due to the imperfect optics and finite collection efficiency and 8 to 12~dB was due to atmospheric loss. The geometric attenuation due to the beam spreading wider than the aperture of the receiver telescope was between 19 to 33~dB, corresponding to a far-field spot size of between 3.5 and 17.9~m, depending on weather conditions. With a tracking accuracy of 3.5~$\mu$rad (0.34~m at the receiver), we had stable count rates for single photons.
We obtained 1171 coincidences during an effective time of 14400~s.
The average channel attenuation was about 44~dB, and the time accuracy was better than 1~ns. We selected linear polarization states
$|H\rangle$, $|V\rangle$ and $|\pm\rangle=(|H\rangle\pm|V\rangle)/\sqrt{2}$ , circular polarization states  $|R\rangle=(|H\rangle+i|V\rangle)/\sqrt{2}$ and $|L\rangle=(|H\rangle-i|V\rangle)/\sqrt{2}$ as the initial states to be teleported. The final data results are shown in Table~\ref{Tab:100kmtele:Result0}.
The experimental results for teleportation fidelity for different initial states range from $76\%$ to $89\%$, with an overall average fidelity of $80\%$. The fidelities for the six teleported states were all well beyond the classical limit of $2/3$. 
%As the weather was keep changing, the coincidence count rates of each state were also slightly different. But in the experiment we measured two orthogonal states simultaneously, so finally we succeeded to eliminate the effect on impact of fidelity leading to the atmospheric shaking.

\section*{Two-link entanglement distribution}

\begin{figure*}[t!]
\begin{center}
\includegraphics[width=0.9\linewidth]{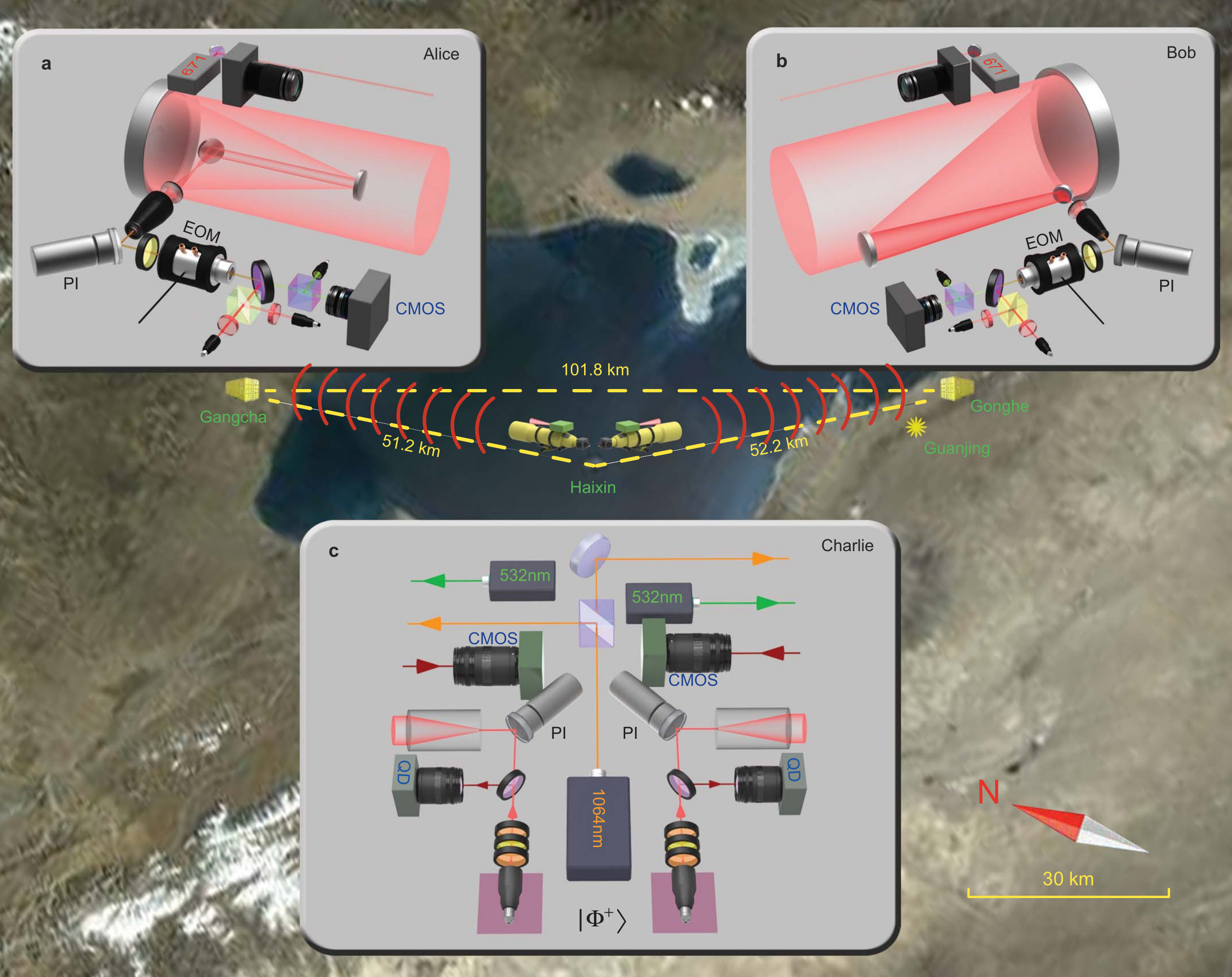}
\vspace{-0.5cm}
\caption{Illustration of the experimental setup for entanglement distribution. \textbf{a},  Alice collects the photon sent by Charlie using a 600 mm Cassegrain telescope. \textbf{b} At the receiver Bob, the photon beams are collected by a 400mm off-axis reflecting telescope. \textbf{c} An entangled photon pair ($\vert \Phi^{+}\rangle$) is created by Charlie at the centre island of Qinghai Lake (Haixin).  Each photons is then guided to a telescope mounted on a two-dimensional rotatable platform.  Then Charlie distributes the two entangled photon to Alice and Bob. The same APT system as in the teleportation experiment is used to create the two-link quantum channel - between Alice and Charlie and between Bob and Charlie.
After receiving the photons, Alice and Bob guide the photons to the detection module by an optical system. This module consists of an EOM, a PBS and two fiber coupled SPCMs. %
\label{Fig_twolink}}
\end{center}
\end{figure*}

In the teleportation experiment, Alice and Charlie are close to each other.
A more common situation would be that Alice is also far away from Charlie. In this case, distribution of entanglement between Alice and Bob is a prerequisite towards quantum teleportation. A feasible solution is to distribute the entanglement to Alice and Bob by Charlie via a two-link channel. To demonstrate the two-link entanglement distribution, we move the entanglement source close to the middle of the free-space channel, an island in the middle of the Qinghai Lake (36$^{\circ}$51$^{\prime}$38.75$^{\prime\prime}$ N, 100$^{\circ}$8$^{\prime}$15.22$^{\prime\prime}$ E), as shown in Fig.~4. To show the novelty of entanglement distribution, Bob move his receiving platform to a local Tibetan family (GongHe 36$^{\circ}$32$^{\prime}$20.66$^{\prime\prime}$N, 100$^{\circ}$33$^{\prime}$45.38$^{\prime\prime}$E) next to  the GuanJing, where it is not possible for Bob to see Alice directly. 
Charlie first prepares the entangled photon pairs in state $\vert \Phi^+\rangle$. Then the entangled photon pairs are sent to Alice and Bob via two telescopes each mounted on a two-dimensional rotatable platform. The distance between Charlie and two receivers is 51.2~km (Alice) and 52.2~km (Bob), and the distance between Alice and Bob is 101.8~km.

The same APT system as in the teleportation experiment is used between Alice and 
Charlie, as well as between Bob and Charlie. Entangled photons are collected by 
telescopes on both sides. In contrast to the 400mm off-axis reflecting telescope used on Bob's side, Alice uses a 600 mm Cassegrain telescope to collect the photons.
In order to confirm the successful entanglement distribution between the two receivers, we measure the $S$ parameter in the Clauser-Horne-Shimony-Holt (CHSH) type Bell's inequality\cite{CHSH}:
$
S = \left| {E\left( {\varphi _A ,\varphi _B } \right) - E\left( {\varphi _A ,\varphi '_B } \right) - E\left( {\varphi '_A ,\varphi _B } \right) - E\left( {\varphi '_A ,\varphi '_B } \right)} \right|,
$
where $E\left( {\varphi _A ,\varphi _B } \right)$ is the correlation function, and $\varphi _A$ and $\varphi '_A$ ($\varphi _B$ and $\varphi '_B $)
are the measured polarization bases of the photon in Alice's (Bob's) hand. In the measurement,
the polarization settings are (0, $\pi$/8), (0, 3$\pi$/8), ($\pi$/4,
$\pi$/8) and ($\pi$/4, 3$\pi$/8) and randomly selected. Each 
measuring module consists of a HWP, a fast electro optical modulator (EOM) and its modulation logical circuit, a PBS, and two fiber-coupled SPCMs.  The optical axes of the EOM are set at 22.5$^{\circ}$ to act as HWPs which transform the diagonal (antidiagonal) polarization into a horizontal (vertical) polarization and viceversa when half-wave voltages are applied. The EOMs act as absent wave plates when zero-wave voltages are imposed. 
Quantum random number generators (QRNG) are used to produce the random digital series between zero-wave and half-wave voltage. Together with the HWP, which is set at 0$^\circ$ at Alice's side and 11.25$^\circ$ at Bob's side, and the QRNG, the EOM randomly switches between the two desired measurement bases - 0 and $\pi$/4 for Alice and $\pi$/8 and 3$\pi$/8 for Bob. 
%The raw bit rate of the QRND is 4 Mbit per second. In our setup, 66.7 kHz random numbers are compressed from the QRNGs in the logical circuit, and then high voltage pulses are generated to switch the voltages between zero and half wave voltages according to the number series. 

\begin{figure}[t!]
\begin{center}
\includegraphics[width=0.8\linewidth]{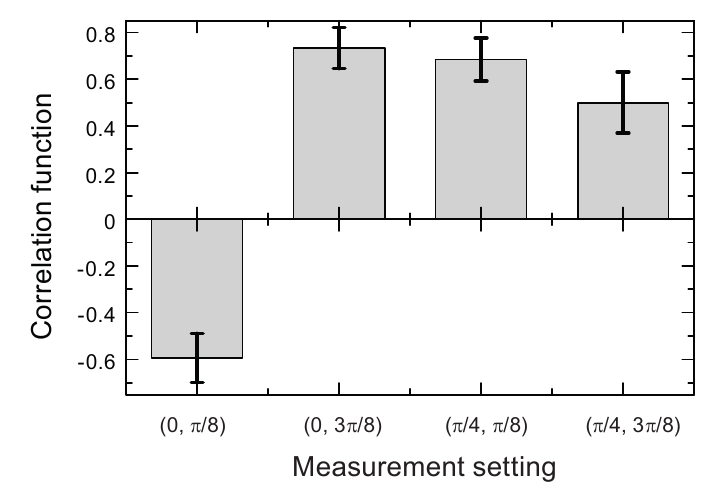}
\vspace{-0.5cm}
\caption{Correlation functions of a CHSH-type Bell's inequality for entanglement distribution. The measurement setting $\left( {\varphi _A ,\varphi _B }\right )$  represents  the measured polarization bases of photons by Alice and Bob, respectively. Error bars represent statistical errors, which are $\pm$1 s.d.
\label{Fig_Bell}}
\end{center}
\end{figure}

Finally, we obtained 208 coincidences during an effective time of 32000~s. Comparing with the counts of our entanglement source, the channel attenuation varied from 66~dB to 85~dB with an average value of 79.5~dB.
For 20~cm aperture satellite optics at an orbit height of 600km and 1~m aperture receiving optics, the total loss for a two-downlink channel between a satellite and two grounds stations is typically about 75~dB.
%was about 80~dB. The effective data show that depending on weather conditions, this attenuation was slowly shaking between 66dB and 85dB. Not that, typically for a two-downlink channel of the satellite to two ground stations,  the total loss is about 75 dB (calculated for 20~cm satellite optics at an orbit height of 500~km). 
The measured correlation functions (shown in Fig.~\ref{Fig_Bell}) resulted in$S=2.51\pm0.21$, which violates Bell's inequality by 2.4 standard deviations.

In addition, our experiment closed the locality loophole. The entangled photon pairs were distributed along two opposite directions to Alice and Bob. 
The distance between Alice and Bob is 101.8~km, which takes 340~$\mu$s for light to traverse, and the path difference between Charlie and Alice and between Charlie and Bob is 1~km, which results in a 3~$\mu$s delay between the two measurement events.  
%The distance between Alice and Bob is 97km ($\sim$324~$\mu$s for light), while the difference of the two distances--from the Charlie to Alice and from the Charlie to Bob--is 1~km (3~$\mu$s delay between two measurement events). 
Thus, Alice and Bob are space-like separated. Furthermore, the two receivers used fast EOMs to switch between the two possible polarization bases. The two EOMs were controlled by two independent QRNGs, each of which generates a random number every 20~$\mu$s (less than 340~$\mu$s). Thus the measurement setting choices are also space-like separated. Hence, the locality loophole is closed. 
%On the other hand, the QRNGs were placed at the two receivers and an additional delay of 3~$\mu$. If the hidden variables were generated with the entangled photon pairs, they need 173~$\mu$s (170~$\mu$s (photon from the source to receiver) +3~$\mu$s (time for random number to control the EOMs)) to affect the measurement. But it is only 170.06~$\mu$s (170~$\mu$s (photon from the source to receiver) +30~ns (photon run in fiber which placed in Transmitter) +30~ns (photon run in the telescope in the receiver)) when the photon passed through the EOMs. This means that the photon had been modulated by the EOMs when the hidden variables affected the EOMs. So our experiment closed the  freedom-of-choice loophole.

\section*{Discussion}

In this work, based on multi-photon entanglement, we experimentally realized free-space quantum teleportation for an independent quibit over a 35--53~dB loss one-link channel. In comparison with previous multi-photon experiments\cite{Marcikic03,Ursin04}, we have enhanced the transmission distance by two orders of magnitude to 97~km. 
Furthermore, we demonstrated the successful distribution of an entangled photon pair over a two-link free-space optical channel to two receivers separated by more than 100~km.
In contrast to previous long-distance free-space experiments with an entangled photon pair using only one-link channels\cite{Ursin07,Fedrizzi09}, our two-link experiment requires tracking and synchronization with three different locations. 
Our two-link experiment, which most comparable with satellite-to-ground quantum entanglement distribution, has achieved a distance between two receivers by an order of magnitude larger than in all previous experiments. 
This shows the feasibility of achieving  two-link quantum teleportation with either the original scheme\cite{Bennett93, Bouwmeester97} or the modified scheme\cite{Boschi98,Jin10}.
Our results show that even with a high-loss ground-to-satellite uplink channel, or satellite-to-ground two-downlink channel, quantum teleportation and entanglement distribution can be realized. Furthermore, our APT system can be used to track an arbitrarily moving object with high frequency and high accuracy, which is essential for future satellite-based ultra-long-distance QC. We hope our experiment will boost the tests of the quantum fundations on a global scale.

This work has been supported by the NNSF of China, the CAS, the National Fundamental Research Program 
(under Grant No.  2011CB921300).

\section*{Appendix}

\renewcommand{\thefigure}{A\arabic{figure}}
 \setcounter{figure}{0}
\renewcommand{\theequation}{A.\arabic{equation}}
 \setcounter{equation}{0}
 \renewcommand{\thesection}{A.\Roman{section}}
\setcounter{section}{0}

\section{Acquiring, pointing and tracking system}

Considering a total optical channel of almost 100~km at near-surface atmosphere, one of the most crucial points is that how to guarantee the stability of the efficiency of the quantum channel. Generally speaking, there are mainly two factors depraving the stability of the quantum channel. One is the instabilities of transmitting platforms mainly caused by ground settlement and mechanical stress relieving, which induce quantum-signal beam deviation slowly and drastically. The other one is angle-of-arrival (AOA) fluctuations of quantum-signal beam caused by atmospheric turbulence, which is at frequency of less than 100~Hz. For solving above problems, we equip an acquiring, pointing and tracking (APT) system on both transmitting and receiving sides.

\subsection{APT system for quantum teleportation}

\begin{figure}
\begin{center}
\includegraphics[width=0.95\linewidth]{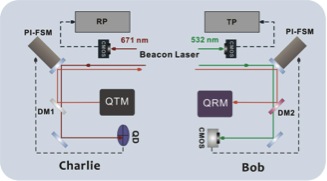}
\caption{\label{APT_tp}\textbf{A sketch of the tracking systems equipped on Charlie's and Bob's station.} DM1: dichroic mirror (T: 671 nm; R: 788 nm); DM2: dichroic mirror (T: 532 nm; R 788 nm); QD: four-quadrant detector; QTM: quantum transmitting module (see Fig. 2 in the main text for detail); QRM: quantum receiving module (see Fig. 2 in the main text for details); RP: the two-dimensional rotatable platform; TP: telescopeÕs rack.}
\end{center}
\end{figure}

On the sender Charlie's side, as shown in Fig.~\ref{APT_tp}, the APT is composed of a two-stage system, namely coarse tracking and fine tracking respectively. For the coarse tracking system, the actuator is a two-dimensional rotatable platform in both direction in the azimuth and pitching (blue arrows in Fig. 2a in the main text). A wide-angle complementary metal oxide semiconductor (CMOS) camera is used to detect the beacon laser coming from the receiver Bob to provide a feedback loop for coarse tracking. Considering the main function of tracking system is to perform the coarse alignment with the receiver and offset the disturbance due to ground settlement and mechanical deformation, the coarse tracking is designed with 30~mrad of tracking field-of-view (FOV), 30~$\mu$rad of tracking accuracy and 10~Hz of closed-loop bandwidth. Furthermore, for fine pointing to the receiver and restraining the beam quiver due to atmospheric turbulence, we implement the fine tracking exploiting a four-quadrant detector (QD) for providing a feedback loop, a fast steering mirror (FSM) driven by piezo ceramics for an actuator. Different with the coarse tracking, the fine tracking system shares the same optical path as the quantum channel and is later separated by a dichroic mirror (DM) (see Fig.~\ref{APT_tp} and Fig. 2a in the main text). Thus, a much higher tracking accuracy (3~$\mu$rad) can be obtained. The response frequency of the QDs is 100 kHz. With the optimal control software with proportion-integration-differentiation (PID) arithmetic, the closed-loop bandwidth of the fine tracking is more than 150~Hz (see Fig.~3 in the main text), which is sufficient to overcome most of the atmospheric turbulence \cite{Strohbehn78}. Finally, with above designed system we get the tracking accuracy better than 3.5~$\mu$rad over 97 km free-space link.

At the receiver Bob's side, there are also coarse and fine tracking systems, by close-loop control via the telescope's own rack and piezo ceramics. Since the main purpose of the tracking system at the receiver is to reduce the low frequency shaking due to ground settlement and passing vehicles, the closed-loop bandwidth is about 10~Hz.

The brief introduction on the procedure of establishing the free-space quantum optical link is as follows: First, Charlie achieved acquiring by Global Positioning System (GPS) coordinates and light guide of Bob; Then, a continuous green laser (532~nm, 200~mW, 1.5~mrad) coaxial with the entangled photon 3, is sent from Charlie to Bob; Third, when Bob acquires and tracks the green laser, he fires a high-power beacon laser (671~nm, 2~W, 200~$\mu$rad) to Charlie for her tracking system; Fourth, utilizing the beacon laser coming from Bob, Charlie turns on her APT system; Finally, after activating the APT systems of both sides, Charlie transmits the entangled photon 3 and a pulsed synchronization laser (1064~nm, 10~kHz, 50~mW, 200~$\mu$rad) to Bob. The APT system is designed for tracking an arbitrarily moving object, which can be directly utilized for a satellite-based quantum communication experiment. In experiments between fixed locations, like the experiment presented here, the first two steps, acquiring and pointing, do not need to be done every day.

\subsection{APT system for entanglement distribution}

\begin{figure}
\begin{center}
\includegraphics[width=0.95\linewidth]{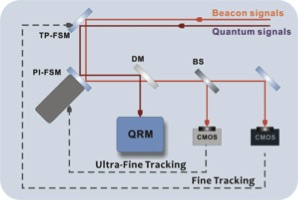}
\caption{\label{APT_ED}\textbf{A sketch of fine tracking system equipped on Alice's telescope in QED experiment}. DM: dichroic mirror; BS: beam splitter; QAM: quantum receiver module (see the details in Fig. 4a in the main text).
}
\end{center}
\end{figure}

For the experiment of long-distance quantum entanglement distribution (QED) over two-link free-space channel, the APT systems on Charlie's (the transmitter) and Bob's (the receiver at Gonghe) station are almost identical with the ones used in the teleportation experiment. To obtain a more efficient and stable free-space optical channel, we employ a larger optical telescope with 600 mm diameter and a high-speed APT system at Alice's (the other receiver) station.

In order to solve the problems of dithering of beam light caused by the low order effects of atmospheric turbulence, we have made great efforts on improving the existing APT of Alice's receiver telescope. The main improvement is that we divide the original fine tracking system into two levels, called fine tracking and ultra-fine tracking (see Fig.~\ref{APT_ED}). For the fine tracking, a CMOS camera with 400~Hz of sampling rate and 15um of per pixel size is used. An indigenous FSM with a shrapnel structure, namely TP-FSM is used as an actuator in the fine tracking system. The overall closed-loop bandwidth is about 15~Hz. Considering the equivalent focal length is 4.3m, the fine tracking system provide a relatively large FOV. For the ultra-fine tracking, a higher sampling rate (2~kHz) CMOS camera is exploied. And due to the longer equivalent focal lengths (7m) and a higher performance actuator (FSM based on piezo, namely PI-FSM), the ultra-fine tracking provide a higher tracking accuracy (2~$\mu$rad) and closed-loop band-with (75~Hz). But since the very narrow FOV of the ultra-fine tracking, we should run the tracking system in combination mode in which the two level tracking system are both turned on to obtain the best tracking effect.

\iffalse
In terms of the high order effects of atmosheric turbulence, such as the beam light spread, an adaptive optics (AO) system is requisite. However, due to significantly enhancement of Charlie-to-Alice's channel efficiency, the total channel efficiency can satisfy the experimental requirements basically. Therefore, the AO system is not 
\fi

Beside the tip tilt error due to the turbulence, there is also higher order effect by the turbulence, which expressed as beam spread. As the telescope aperture we used at the receiver is much smaller than the far-field spot size, any speckle effects results more in geometric attenuation than efficiency fluctuation, which cannot be directly observed with our system. To narrow the far field spot, one needs to increase the tracking bandwidth to more than kHz with new technologies, for example an adaptive optics (AO) system.

\section{synchronization accuracy}

In the experiment, as shown in Fig.~2 and Fig.~4 in the main text, a pulsed infrared laser (1064~nm, 10~kHz, 50~mw, 200~$\mu$rad) is used for synchronization. The pulse length of the synchronization laser is 2.65~ns (FWHM), with the rising edge of 2~ns. A wireless Bridge is used for data transmission and classical communication between Alice and Bob.  A pulse per second by the GPS are added to synchronize the starting time. 

First we collect the laser with multi-mode fibers (200~$\mu$m, NA~0.22) and then guide the light into fast photoreceivers at both receiving sides. Then we deal with the forefront of detected signals with constant fraction discriminator (CFD) technique and utilize a high-accuracy Time-to-Digital Converter (TDC) with 100~ps time resolution precision to record the arrival time of the synchronization signals at both Alice's and Bob's station.

Through the above methods, we had overcome most of the time jitter caused by the laser pulse itself and energy shaking. 
In the experiment, the quantum signal was collected by SPCM and an additional time  jitter of 350~ns has to be considered. 
Finally we achieved time synchronization for the quantum channel better than 1~ns over 100km distance (See Fig.~\ref{synchron})

%A high-accuracy Time-to-Digital Converter is used to  independently record the arrival time of the 1064~nm laser signals at both Alice's and Bob's station.

\begin{figure}
\begin{center}
\includegraphics[width=0.95\linewidth]{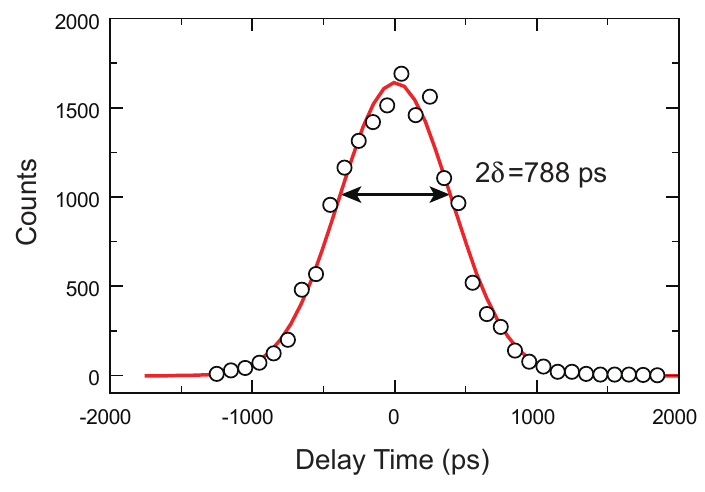}
\caption{\label{synchron}\textbf{A typical recorded arrival time of the 1064~nm laser signals at both stations.} Synchronization accuracy of 2$\delta$=788(13)~ps was observed, $\delta$ refers to the width of the gaussian fit (solid line).
\label{synchron}}
\end{center}
\end{figure}

\end{document}